\documentclass[prl,aps,twocolumn]{revtex4-1}
\usepackage{amsmath}
\usepackage{amssymb}
\usepackage{amsthm}
\usepackage{bbm} 
\usepackage{mathrsfs}
\usepackage{bm}
\usepackage{graphicx,color}

\newcommand{ \sD }{ \mathcal{D} }
\newcommand{ \sA }{\mathcal{A}}
\newcommand{ \sL }{\mathcal{L}}

\newcommand{ \sFidelity }{\mathcal{F}}

\newcommand{\Tr}{\mbox{Tr}}                                        
\newcommand{\nn}{\nonumber}

\begin{document}

\title{Ignorance Is Bliss: General and Robust Cancellation of Decoherence via No-Knowledge Quantum Feedback}

%Some alternative titles (with different combinations thereof)
%\title{(Quantum) Ignorance is bliss: useful quantum feedback without knowledge}
%\title{Complete cancellation of decoherence without knowledge}
%\title{Measuring nothing cancels everything: useful quantum feedback without knowledge}
%\title{Measuring nothing cancels everything: Complete cancellation of decoherence via a no-knowledge measurement feedback protocol}

%From Michael
%title{Robust quantum feedback without knowledge}
%My favourite composite title
%\title{Ignorance is bliss: General and robust cancellation of decoherence via no-knowledge quantum feedback}}

%I feel like we should emphasise the generalness as well. I know quantum feedback is a silly word. But perhaps we don't want to limit ourselves to measurement feedback, as there is no particular reason to think this can't be done coherently.

\author{Stuart S. Szigeti$^1$}
\author{Andre R. R. Carvalho$^{2,3}$}
\author{James G. Morley$^4$}
\author{Michael R. Hush$^4$}

\affiliation{$^1$ARC Centre for Engineered Quantum Systems, The University of Queensland, Brisbane, QLD 4072, Australia}
\affiliation{$^2$Department of Quantum Science, Research School of Physics and Engineering, The Australian National University, Canberra, ACT 0200, Australia}
\affiliation{$^3$ARC Centre for Quantum Computation and Communication Technology, The Australian National University, Canberra, ACT 0200, Australia}
\affiliation{$^4$School of Physics and Astronomy, University of Nottingham, Nottingham, NG7 2RD, United Kingdom}

\date{\today}

\begin{abstract}

A ``no-knowledge'' measurement of an open quantum system yields no information about \emph{any} system observable; it only returns noise input from the environment. Surprisingly, performing such a no-knowledge measurement can be advantageous. We prove that a system undergoing no-knowledge monitoring has reversible noise, which can be cancelled by directly feeding back the measurement signal. We show how no-knowledge feedback control can be used to cancel decoherence in an \emph{arbitrary} quantum system coupled to a Markovian reservoir that is being monitored. Since no-knowledge feedback does not depend on the system state or Hamiltonian, such decoherence cancellation is guaranteed to be general, robust and can operate in conjunction with any other quantum control protocol. As an application, we show that no-knowledge feedback could be used to improve the performance of dissipative quantum computers subjected to local loss.

\end{abstract}

\pacs{03.65.Yz, 42.50.Lc, 03.65.Ta, 03.67.Pp, 03.67.-a}

\maketitle

\noindent
``More signal, less noise'' is the guiding philosophy of experimental science. Increasing measurement sensitivity is a proven strategy for pushing the frontiers of science and technology, yielding improved knowledge and control over Nature. However, at the quantum scale physics pushes \emph{back} by imposing a fundamental limit on the signal-to-noise ratio by virtue of Heisenberg's Uncertainty Principle~\cite{Dowling:1998, Clerk:2010}. Nevertheless, ``more signal, less noise'' also guides the design of protocols for the measurement and control of quantum systems, such as squeezed state photon \cite{Scully:1997} and atom \cite{Robins:2013} interferometry, optimal parameter estimation \cite{Wiseman:2010}, weak measurement \cite{Braginsky:1996}, measurement-based feedback control~\cite{Handel:2005, Wiseman:2010} and adaptive measurement \cite{Wiseman:2009}. In this Letter, we take the unorthodox ``no signal, only noise'' approach, and consider measurements that are pure noise, and therefore give \emph{no knowledge of the quantum state whatsoever}. From a quantum control perspective, one intuitively expects such \emph{no-knowledge measurements} to be unworthy of study, since robust feedback control requires at least some (and preferably good) knowledge of the system state. On the contrary, we show that a measurement-based feedback protocol based on no-knowledge monitoring can be used to remove decoherence - the bane of quantum technology - from an arbitrary quantum system coupled to a Markovian environment that can be monitored. 

Although the ``no signal, only noise'' approach is unorthodox, it has been considered within the context of channel correction. In \cite{Gregoratti:2003, Buscemi:2005, Buscemi:2007a}, it was proven that coherence could be recovered in a noisy channel provided the conditional evolution was random unitary. Consequently, complete correction is in principle possible for systems with dimension $d \leq 3$. Furthermore, it was proven that measurements that returned a small amount of knowledge (``little signal, mostly noise'') provided a good error correction strategy, and a trade-off relation between information extraction and correction efficacy was established \cite{Buscemi:2007}. Our no-knowledge feedback scheme is consistent with these results, however it goes several steps further as 1) it concretely shows how decoherence can be cancelled in a system of arbitrary dimension, with arbitrary coupling to a Markovian environment, and 2) it provides the explicit physical description of both the measurement and the conditional evolution via our use of the continuous quantum measurement framework.

Attempts to mitigate decoherence have resulted in significant successes, including the development of error correction codes \cite{Shor:1995, Calderbank:1996, Ekert:1996, Gottesman:1996}, dynamical decoupling \cite{Viola:1999}, reservoir engineering \cite{Poyatos:1996, Carvalho:2001}, feedback control \cite{Doherty:2001, Ganesan:2007, Szigeti:2009, Szigeti:2010, Hush:2013}, and the engineering of decoherence-free subspaces \cite{Zanardi:1997, Lidar:1998}. Nevertheless, decoherence has yet to be adequately tamed. In our proposal, decoherence is cancelled by directly feeding the no-knowledge measurement signal back into the system, in effect turning quantum noise against itself. The scheme only requires knowledge of the decoherence channel to be cancelled; no knowledge of the system state is required. It is consequently effective, robust and can be used in conjunction with other quantum control protocols. This demonstrates that meaningful feedback control without knowledge is not only possible, but desirable. 
 %Unlike many other methods of decoherence reduction, this feedback operates with no \emph{a priori} knowledge of the initial state of the system and is consequently effective, robust and can be used in conjunction with other quantum control protocols. This demonstrates that meaningful feedback control without knowledge is not only possible, but desirable. 

{\bf No-knowledge measurements.}
Consider a system with Hamiltonian $H$ that interacts with a Markovian reservoir via the coupling operator $L$. The system density operator, $\varrho_t$, evolves according to the master equation (ME)
\begin{equation}
	\partial_t \varrho_t = -i[H,\varrho_t] + \sD[L] \varrho_t \equiv \sL \varrho_t, \label{eqn:uncond_ME}
\end{equation}
where $\partial_t \equiv d/dt$, $\sD[Z] \varrho_t = Z \varrho_t Z^\dag - (Z^\dag Z \varrho_t + \varrho_t Z^\dag Z)/2$, and we have set $\hbar = 1$. In principle, it is always possible to indirectly extract information about the system with a projective measurement on the reservoir. In particular, for a homodyne measurement of the environment at angle $\theta$, the conditional system dynamics are described by the Stratonovich stochastic ME (SME) \cite{Wiseman:1993, Wiseman:1994, Wiseman:2010}
\begin{align}
	\partial_t \rho_t 	&= \sL \rho_t  + \sqrt{\eta} \sA[L e^{i\theta}] \rho_t \,  y_\theta(t) - \frac{\eta}{2}\sA^2[L e^{i\theta}] \rho_t,   \label{eqn:system}
\end{align}
where $\rho_t$ is the unnormalized conditional density operator for the system, $\eta$ is the detection efficiency, $\sA[Z] \rho_t = Z \rho_t+ \rho_t Z^\dag$, and $\sA^2[Z]\rho_t= Z (\sA[Z] \rho_t) + (\sA[Z] \rho_t) Z^\dag$. Conditional expectations of system operators are calculated using $\langle X \rangle_t = \Tr[X {\rho}_t]/ \Tr[{\rho}_t]$. The first term of Eq.~(\ref{eqn:system}) corresponds to the unconditional Lindblad ME~(\ref{eqn:uncond_ME}), and gives the unitary dynamics due to the system Hamiltonian and the decoherence caused by the system-reservoir coupling. The second term is the \emph{innovations}, which conditions the system dynamics on the homodyne measurement photocurrent  
\begin{equation}
	y_\theta(t) = \sqrt{\eta} \, \langle L e^{i\theta} + L^\dag e^{-i\theta} \rangle_t + \xi(t),
\label{eqn:meas_signal}
\end{equation}
where $\xi(t)$ is a Stratonovich stochastic integral \cite{Wong:1965, Gardiner:2004}. The final term of Eq.~(\ref{eqn:system}) is the Stratonovich correction (see Supplemental material). Equation~(\ref{eqn:uncond_ME}) is obtained by averaging Eq.~(\ref{eqn:system}) over different realizations of the measurement record, up to a normalization factor.

Equation~(\ref{eqn:meas_signal}) shows that the measurement signal is composed of two parts; the first term represents the knowledge obtained about the system from the measurement, whereas the second term is the corrupting quantum (white) noise input from the reservoir. However, there exist choices of $L$ for which the measurement returns no information about the system operators, which we term a \emph{no-knowledge measurement}. Specifically, when $L$ is Hermitian, homodyne detection of the reservoir at angle $\theta = \pi/2$ is a no-knowledge measurement, since the measurement signal $y_{\pi/2}(t) = \xi(t)$ returns only noise. No-knowledge monitoring appears in early works on continuous quantum measurement as a means of obtaining simpler linear SMEs \cite{Diosi:1988, Milburn:1996}, in the investigation of the localization properties of conditioned states~\cite{Rigo:1997b}, and in the discussion of state estimation~\cite{Doherty:1999b, Doherty:1999c}.
    
\begin{figure}[htb]
\includegraphics[width=\columnwidth]{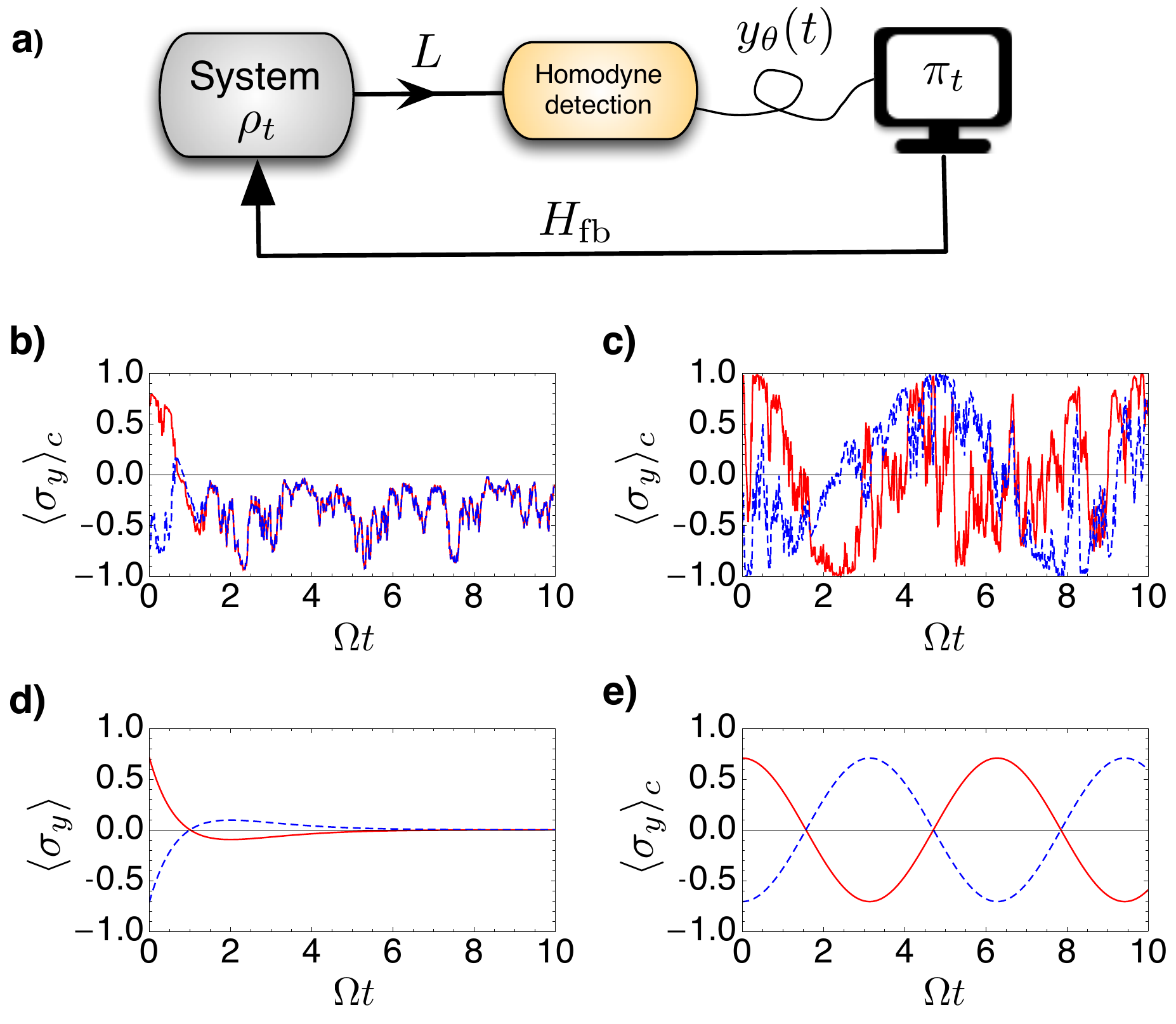}
\caption{(a) Schematic for a measurement-based feedback control protocol. In general, information about the system $\rho_t$ is extracted by monitoring the decoherence channel $L$. The optimal estimate $\pi_t$ is conditioned on the resulting measurement signal $y_\theta(t)$. The system is then controlled with some feedback Hamiltonian $H_\text{fb}$. For our no-knowledge feedback protocol, the feedback is simply a modulation of the no-knowledge measurement signal. (b-e) Particular example of a driven qubit undergoing dephasing with $H= \Omega \sigma_x$,  $L=\sqrt{\gamma}\sigma_z$ and $\Omega/\gamma = 1$. (b, c) Conditional trajectories for $\langle \sigma_y \rangle_c$ when the channel is being monitored (no feedback) with perfect homodyne detection at angles $\theta=4\pi/5$ and $\pi/2$, respectively. Solid red lines represent the dynamics starting from the underlying initial state $\rho_0 = [I + (\sigma_x + \sigma_y)/\sqrt{2}]/2$, while dashed blue lines represent the filter evolution from the (incorrect) initial estimate $\pi_0 = [I + (\sigma_x - \sigma_y)/\sqrt{2}]/2$. Although the estimate $\pi_t$ converges to $\rho_t$ in (b), in the no-knowledge case (c) $\rho_t$ and $\pi_t$ never converge. (d) Dephasing effect for the unmonitored system [\emph{c.f.} Eq.~(\ref{eqn:uncond_ME})]. (e) Dephasing is cancelled by directly feeding back the no-knowledge measurement via the Hamiltonian $H= \Omega \sigma_x + \sqrt{\gamma} \sigma_z y_{\pi/2}(t)$. Despite the filter's inaccurate estimate of $\rho_t$, decoherence is completely removed, demonstrating that accurate knowledge of the system is not required for effective decoherence cancellation.} 
\label{fig:filtervsactual}
\end{figure}

We can examine the effect of a no-knowledge measurement by comparing the evolution of the underlying system state, $\rho_t$, to that of the quantum filter \cite{Handel:2005, Bouten:2007}, $\pi_t$, which is the optimal Bayesian estimate of the system state conditioned on the measurement record (see Supplemental material). The unnormalized quantum filter ${\pi}_t$ evolves according to \cite{Amini:2011, Szigeti:2013} 
\begin{align}
	\partial_t \pi_t 	&= \sL \pi_t  + \sqrt{\eta} \sA[L e^{i\theta}] \pi_t \, y_\theta(t) - \frac{\eta}{2}\sA^2[L e^{i\theta}] \pi_t. \label{eqn:optimalfilter}
\end{align}
Suppose that we have the situation shown in Fig.~\ref{fig:filtervsactual}(a) (without the feedback) where the system is prepared in the state $\rho_0$ and evolves according to Eq.~(\ref{eqn:system}), while an observer, ignorant of the underlying system state, models the system by Eq.~(\ref{eqn:optimalfilter}) with $\pi_0 \not=\rho_0$. In general, information about the system is extracted from the measurement signal and used to update the observer's estimate. This leads to a better estimate of the system state over time, and $\pi_t$ converges to $\rho_t$ in finite time [Fig.~\ref{fig:filtervsactual}(b)]. This is \emph{not} true for a no-knowledge measurement, since the filter is conditioned only on noise. Then Eqs~(\ref{eqn:system}) and (\ref{eqn:optimalfilter}) decouple, and the filter never converges to the system state (see Supplemental material) [Fig.~\ref{fig:filtervsactual}(c)]. 

{\bf Cancelling reservoir noise with no knowledge.}
In classical control theory, a system-observation pair is called unobservable if the initial system state cannot be determined from the measurement signal. A system undergoing a no-knowledge measurement is clearly unobservable, as neither the past or present system state can be determined from the measurement record. One may expect, therefore, that this lack of knowledge renders meaningful measurement-based feedback control impossible. This intuition is incorrect. Although a no-knowledge measurement produces a signal with no dependence on any system observable, the quantum noise that constitutes the signal is precisely the same noise that corrupts the system state. Consequently, by applying an appropriate feedback the no-knowledge measurement signal can be used to cancel the noise corrupting the system's evolution. %thereby removing the deleterious effects of reservoir noise from the system's evolution.
%the noise corrupting the measurement signal is correlated with the reservoir noise coupled to the system. A no-knowledge measurement therefore still contains \emph{information} (in an information-theoretic sense). More interestingly, this information can be used to cancel the deleterious effects of reservoir noise.

Specifically, suppose $L$ is Hermitian, and we make a measurement of the no-knowledge quadrature $\theta = \pi/2$ with perfect efficiency $\eta = 1$. Then Eq.~(\ref{eqn:system}) takes the simple form: 
\begin{equation}
	\partial_t \rho_t = -i\left[H - L \; y_{\pi/2}(t), \rho_t \right]. \label{eqn:reversnoise} 
\end{equation}
Since the dynamics due to the reservoir noise are unitary, their effect is \emph{reversible} and can be entirely cancelled by directly feeding back the measurement signal. Explicitly, by making the replacement $H \to H + L \; y_{\pi/2}(t)$, Eq.~(\ref{eqn:reversnoise}) reduces to $\partial_t \rho_t = -i[H,{\rho}_t]$. 
%Crucially, this feedback only works because the noise that formed the measurement signal was perfectly correlated with the reservoir noise inducing the decoherence. 

What is particularly interesting about no-knowledge feedback is that it works when the system and filter are initially very different [see Figs~\ref{fig:filtervsactual}(d) and (e)]. The reason is that the measurement signal is simply fed back via the Hamiltonian without \emph{any} prior filtering. Indeed, no-knowledge feedback can be successfully implemented with almost no \emph{a priori} knowledge of the underlying system state or dynamics. No-knowledge feedback only requires a correct identification of the no-knowledge quadrature, which depends only on the coupling operator $L$, and the ability to monitor this decoherence channel. \emph{A precise description of the system state and its unitary evolution is not required.} This natural \emph{robustness} \cite{Chen:2013} gives no-knowledge feedback an advantage over other state-dependent methods of decoherence reduction \cite{Vitali:1997}, particularly for systems where the dynamics cannot be precisely quantified.  

When the detection efficiency is imperfect, the effectiveness of no-knowledge feedback is reduced. For the evolution is no longer purely unitary:
\begin{equation}
	\partial_t{\rho}_t = -i[H - \sqrt{\eta} L \; y_{\pi/2}(t), \rho_t] + (1 - \eta) \sD[L]{\rho}_t, \label{eqn:filter_imperfect} 
\end{equation}
and therefore cannot be entirely cancelled by feeding back the measurement signal. Nevertheless, by choosing the no-knowledge feedback $H \to H +\sqrt{\eta} L \; y_{\pi/2}(t)$ the decoherence rate can be reduced by a factor of $(1-\eta)$ [\emph{c.f.} Eq.~(\ref{eqn:uncond_ME})]:
\begin{equation}
	\partial_t \rho_t = -i[H,{\rho}_t] + (1 - \eta)\sD[L] {\rho}_t.
\end{equation}
Experiments with imperfect detection efficiency can therefore still enjoy a significant and robust decoherence reduction by employing no-knowledge feedback.

An analogous result exists for photodetection, where unitary $L$ corresponds to a no-knowledge measurement. Noise is cancelled by applying a unitary gate to the system after the detection of a photon (see Supplemental material).

{\bf Removing decoherence for general $L$.} 
As formulated above, a no-knowledge measurement is only possible when the coupling operator is Hermitian \footnote{Strictly, $L^\dag = L \exp(i \phi)$.}. Since physical observables are Hermitian, direct no-knowledge measurements are possible in many situations. Examples include dephasing in qubits ($L=\sigma_z$) \cite{Nielsen:2010}, optomechanical devices under position measurement ($L=x$) \cite{Doherty:2012} and minimally-destructive detection of Bose-Einstein condensates \cite{Szigeti:2009, Szigeti:2010, Hush:2013, Gajdacz:2013}. However, some common coupling operators, such as the annihilation operator $a$, are not Hermitian. Fortuitously, we can still \emph{remove decoherence for a general} $L$ via a similar measurement-based feedback scheme. Counter-intuitively, this requires an \emph{extra} reservoir with coupling operator $L^\dag$, giving the unconditional dynamics
\begin{equation}
	\partial_t \varrho_t = -i[H,\varrho_t] + \sD[L] \varrho_t+ \sD[L^\dag] \varrho_t. \label{eqn:twosystem}
\end{equation} 
The `trick' is to recognize that $\sD[L]{\rho}_t + \sD[L^\dag]{\rho}_t = \sD[L_+]{\rho}_t + \sD[L_-]{\rho}_t$, where $L_\pm = i^{(1\mp1)/2}(L \pm L^\dag)/\sqrt{2}$ are Hermitian. Thus $L_\pm$ are effective coupling operators that admit no-knowledge measurements. 

\begin{figure}[tb]
\includegraphics[width=\columnwidth]{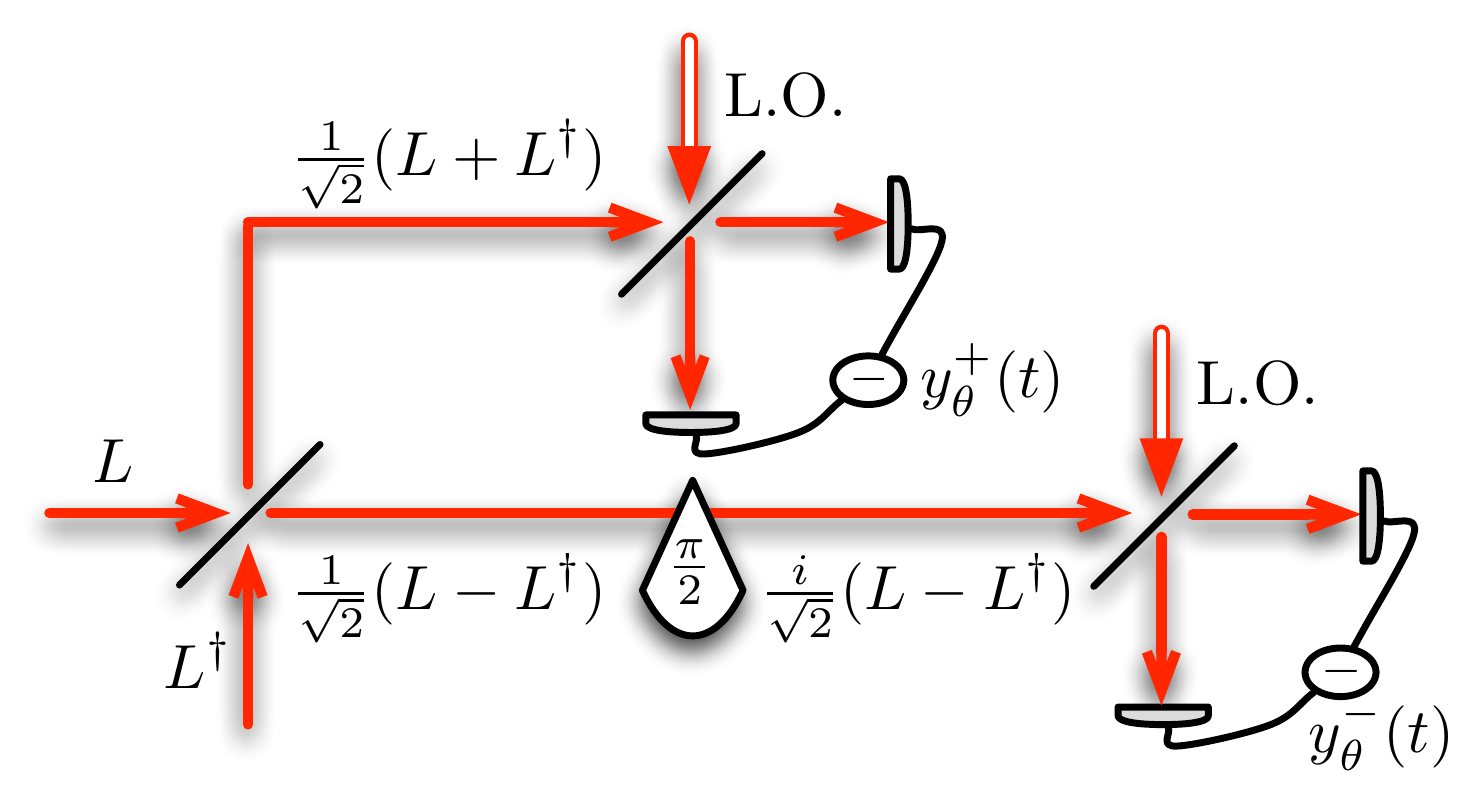}
\caption{Scheme for engineering measurements of $L_\pm$ from the outputs of couplings $L$ and $L^\dag$.}
\label{fig: general_L_setup}
\end{figure}
Measurements of $L_\pm$ are possible by taking the output channels of both reservoirs, mixing them via a 50:50 beamsplitter, introducing a relative phase shift of $\pi/2$ and subsequently measuring each output with homodyne detection (see Fig.~\ref{fig: general_L_setup}). This yields the two measurement signals $y^\pm_\theta(t) = 2\sqrt{\eta} \, \cos \theta \langle L_\pm \rangle_t + \xi_\pm(t)$, where $\xi_\pm(t)$ are independent Stratonovich noises. No-knowledge measurements of $L_\pm$ occur for quadrature angle $\theta = \pi/2$. The beamsplitting step of the feedback protocol is vital, and has no classical analogue, making our result a \emph{quantum} feedback protocol.

The evolution of $\rho_t$ under these no-knowledge measurements is given by a straightforward generalization of Eq.~(\ref{eqn:filter_imperfect}):
\begin{align}
	\partial_t{\rho}_t 	&= -i\left[H - \sqrt{\eta} ( L_+ y^+_{\pi/2}(t) + L_- y^-_{\pi/2}(t)), {\rho}_t\right]  \notag \\
				& + (1-\eta) ( \sD[L]{\rho}_t + \sD[L^\dag]{\rho}_t ). \label{eqn:genreversenoise}
\end{align} 
Finally, we directly feed the measurement signals back via $H \to H + \sqrt{\eta} ( L_+ y^+_{\pi/2}(t) + L_- y^-_{\pi/2}(t))$:
\begin{align}
	\partial_t{\rho}_t 	&= -i[H,{\rho}_t] + (1-\eta) ( \sD[L]{\rho}_t + \sD[L^\dag]{\rho}_t).\label{eqn:gennonoise}
\end{align} 
The original decoherence in the system has been suppressed by the factor $(1-\eta)$, admittedly at the cost of introducing additional decoherence due to $L^\dag$. However, in the perfect detection efficiency limit, $\eta \rightarrow 1$, all decoherence is eradicated from the system. 

The successful implementation of our scheme requires some level of reservoir engineering and monitoring. In principle, such dissipative engineering is possible for a range of physical systems. For example, Carvalho and Santos~\cite{Carvalho:2011} showed how to engineer an additional $\sigma_+$ reservoir to the spontaneous emission decoherence of two qubits. This system is a specific instance of Eq.~(\ref{eqn:genreversenoise}) for $L = \sigma_-$, and allows for the protection of entanglement via the environmental monitoring~\cite{Carvalho:2011,Santos:2011,Vogelsberger:2010, Guevara:2014}, or even quantum computation when applied to multiple qubits~\cite{Santos:2012}. % for the purposes of quantum computation, albeit extended for multiple qubits such that the jumps corresponded to two-qubit entangling gates.
Although neither paper considered the possibility of cancelling decoherence via no-knowledge feedback, implementing such feedback would be straightforward via the inclusion of the feedback Hamiltonian $H =(\sigma_- +\sigma_+)y^+_{\pi/2}(t)/\sqrt{2} + i(\sigma_- -\sigma_+) y^-_{\pi/2}(t)/\sqrt{2}$. This Hamiltonian simply corresponds to the application of two classical fields resonant to the qubit transition and modulated by the measurement signals.  

From an experimental standpoint, the homodyne monitoring and modulated feedback driving should be relatively simple to implement in a variety of physical systems. In particular, specific homodyne quadratures can be chosen with a high degree of accuracy, as is routinely done in tomography, and with high efficiencies. More challenging is the reservoir engineering step and the efficient collection of the decoherence channel, which ultimately limits the overall efficiency $\eta$. Nevertheless, recent demonstrations in systems as diverse as superconducting qubits \cite{Murch:2012, Murch:2013, Shankar:2013}, cavity QED experiments \cite{Gleyzes:2007}, and ion traps \cite{Myatt:2000, Bushev:2006} indicate that an experimental realization of our scheme is entirely plausible in the near future. For example, \cite{Murch:2013} reported $\eta=0.49$ when monitoring a cavity field coupled to a superconducting qubit, and efficiencies above 90\% are achievable via coupling an ancilla to the superconducting qubits \cite{Groen:2013}. In microwave cavity experiments, cavity field monitoring with $\eta = 0.5$ has been demonstrated \cite{Guerlin:2007}. 

{\bf Application: dissipative quantum computing.} It was recently shown that appropriately engineered quasi-local dissipation can be used to perform universal quantum computation (UQC) \cite{Verstraete:2009, Kraus:2008}. Although such dissipative quantum computing (DQC) is robust to decoherence in principle, in practice it is likely to suffer from local errors due to the presence of local loss. For traditional UQC, local errors can be corrected via quantum error correction (QEC) codes. Indeed, the threshold theorem proves that traditional UQC can be scaled to large numbers of qubits, even when local errors are present, provided QEC is in operation \cite{Nielsen:2010}. However, QEC requires precisely timed projective measurement and conditional operations, hence adding this capacity to DQC greatly complicates the engineering of these systems \cite{Kastoryano:2013}.  

\begin{figure}[tb]
\includegraphics[width=\columnwidth]{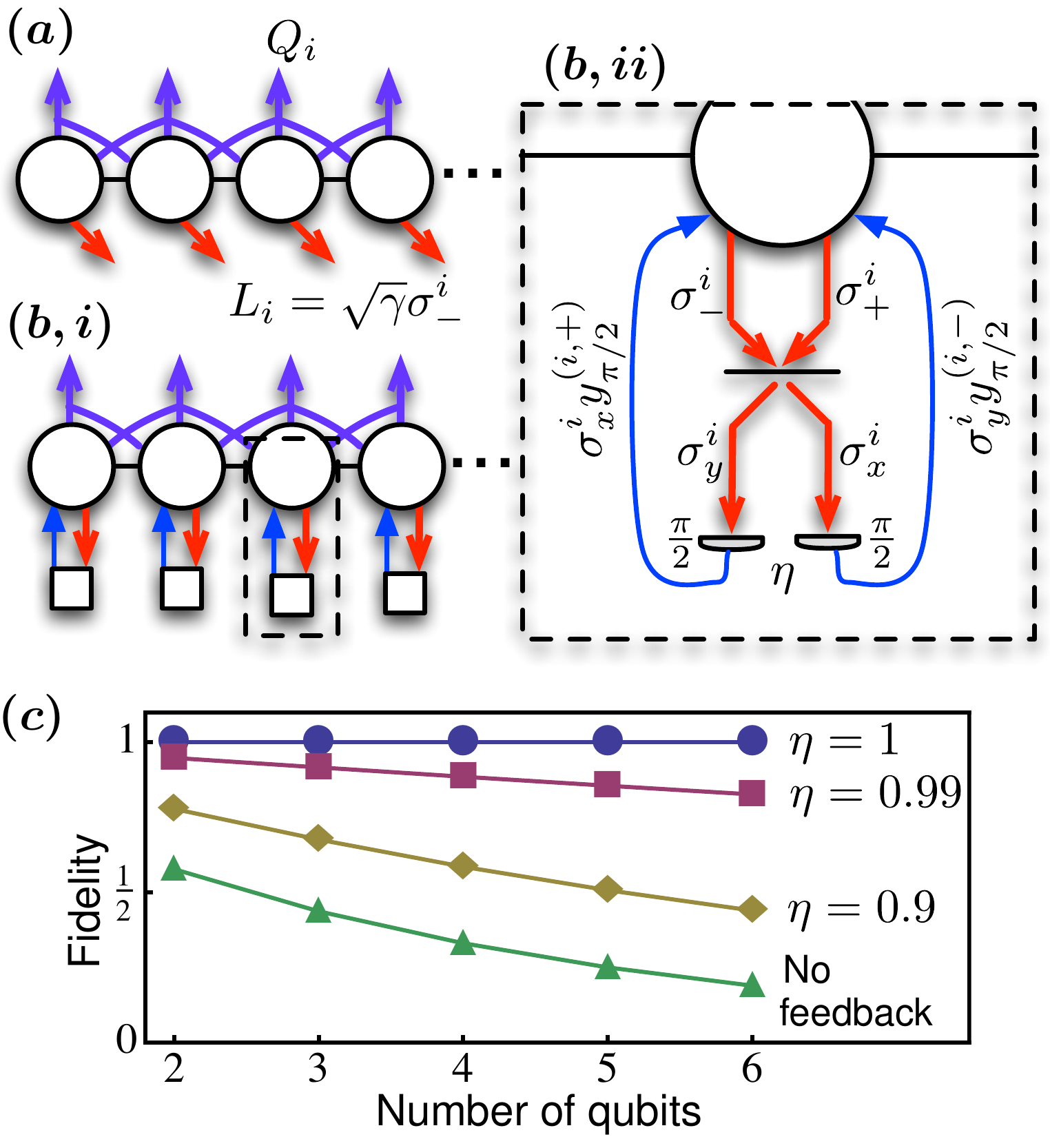}
\caption{(a) A DQC setup with an $N$ qubit chain coupled to quasi-local operators $Q_i$ and local loss operators $L_i$. As demonstrated in (c), a loss rate of $\gamma/\alpha = 10$ decreases the fidelity between the target cluster state and the system steady state (green triangles), and decreases it more severely for a larger number of qubits. (b i) The errors introduced by the local loss are corrected by applying our no-knowledge feedback protocol on each qubit. (b ii) For each qubit a no-knowledge measurement is constructed by coupling an additional reservoir $\sqrt{\gamma} \sigma_+^i$ and measuring $\sigma_x^i$ and $\sigma_y^i$ at a homodyne angle of $\pi/2$ as summarized in Fig.~\ref{fig: general_L_setup}. Decoherence is cancelled by feeding back $H =  \sqrt{\eta \gamma} \sum_i[ \sigma_x^i y^{(i, +)}_{\pi/2}(t) + \sigma_y^i y^{(i, -)}_{\pi/2}(t)]$. (c) The fidelity as a function of system size for no feedback (green triangles), and no-knowledge feedback with detection efficiency $\eta = 0.9$ (yellow diamonds), $\eta = 0.99$ (red squares) and $\eta = 1$ (blue circles).}
\label{fig:dissquant}
\end{figure}

We provide a simpler solution. Provided the cause of the local errors is diagnosable, no-knowledge feedback can be used to remove their effect. Crucially, the feedback will work concurrently with any quantum computation. To show this, we consider the effect of local loss on a DQC algorithm designed to generate a linear cluster state [see Fig.~\ref{fig:dissquant}(a)]. A series of $N$ qubits evolve under the influence of quasi-local dissipators $Q_i =  \sqrt{\alpha} (1+\sigma_z^{i-1} \sigma_x^{i} \sigma_z^{i+1}) \sigma_z^{i}/2$ (with special cases $Q_1 = \sqrt{\alpha} (1+ \sigma_x^{1} \sigma_z^{2}) \sigma_z^{1}/2$ and $Q_N = \sqrt{\alpha} (1+ \sigma_z^{N-1} \sigma_x^{N}) \sigma_z^{N}/2$ at the boundaries) and local loss operators $L_i = \sqrt{\gamma} \sigma_-^i$, such that the ME for the whole system is $\partial_t\varrho_t = \sum_i^N (\sD[Q_i] + \sD[L_i])\varrho_t$. The steady state, $\rho_{ss}$, for the system when there is no local loss ($\gamma = 0$) is a cluster state, $\rho_{ss} = \rho_{\rm cluster}$. However, when local loss is present ($\gamma \not = 0$), the steady state of the system is no longer the target cluster state. As shown in Fig.~\ref{fig:dissquant}(c), the fidelity $\sFidelity = \sqrt{\rm Tr\left[\rho_{ss} \, \rho_{\rm cluster} \right]}$ between the target cluster state and the actual steady state rapidly decreases with system size. However, when no-knowledge feedback is implemented as depicted in Figs~\ref{fig:dissquant}(b), the decline in the fidelity as a function of system size is arrested. Engineering the additional local dissipator $\sigma_+^i$~\cite{Carvalho:2011,Santos:2011} required for this feedback should be trivial in comparison to engineering the quasi-local dissipators $Q_i$. Figure~\ref{fig:dissquant}(c) quantifies the effectiveness of the no-knowledge feedback, demonstrating that the fidelity improves as the detection efficiency increases, with the creation of a perfect cluster state possible when $\eta = 1$. In fact, since no-knowledge feedback can operate concurrently to any DQC algorithm, it could be included \emph{in addition to QEC}. Hence no-knowledge feedback with an imperfect detection efficiency may reduce the error rate to the threshold required for truly scalable DQC.

DQC is just one of many possible quantum technologies that could be improved, or made possible, by the general and robust reduction of decoherence via no-knowledge feedback. However, since no-knowledge feedback can operate in conjunction with other quantum control protocols, it does not compete with other decoherence reduction methods (e.g. QEC), but rather \emph{complements} them. Furthermore, given the simplicity of no-knowledge feedback, we suspect that no-knowledge \emph{coherent} feedback control is a strong possibility. The many advantages of no-knowledge feedback strengthen the case for more reliable and robust dissipation engineering, as this is a vital ingredient for the cancellation of general forms of decoherence.

\textbf{Acknowledgements.} Numerical simulations were performed using XMDS2 \cite{Dennis:2012} on the University of Queensland (UQ) School of Mathematics and Physics computer `obelix', with thanks to Elliott Hilaire and Ian Mortimer for computing support. SSS acknowledges the support of Ian P. McCulloch and the Australian Research Council (ARC) Centre of Excellence for Engineered Quantum Systems (project no. CE110001013). ARRC acknowledges support by the ARC Centre of Excellence for Quantum Computation and Communication Technology (project no. CE110001027). MRH acknowledges support by EPSRC Grant no. EP/I017828/1 and thanks Zibo Miao for fruitful discussions.  

\bibliography{reversible_noise}

\newpage

\section{Supplementary Material}

\setcounter{equation}{0}

\subsection{I. Stochastic master equation in Stratonovich form}

The effect of a no-knowledge measurement is best understood by examining the Stratonovich version of the conditional master equation. However, the literature far more frequently presents conditional master equations with Ito stochastic integrals. Here we present the connection between the two.

To begin, consider an $n \times 1$ vector of stochastic variables $\textbf{x}_t$ that obeys the following linear Ito stochastic differential equation:
\begin{equation}
	d\textbf{x}_t = \textbf{A}(\textbf{x},t) \textbf{x}_t dt + \textbf{B} (\textbf{x},t) \textbf{x}_t \, dw(t), \label{eqn:SDEitoform}
\end{equation}
where $\textbf{A}$ and $\textbf{B}$ are $n \times n$ matrices, and $dw(t)$ is an Ito Wiener increment satisfying $dw(t)dw(t) = dt$. Furthermore, terms with $dw(t)$ average to zero, making Ito SDEs the convenient choice for most analytic work. Although the differential shorthand is convenient, Eq.~(\ref{eqn:SDEitoform}) is strictly interpreted as an integral equation. In order to express Eq.~(\ref{eqn:SDEitoform}) in terms of the Stratonovich noise $\xi(t)$, we must add the deterministic correction $-\textbf{B}^2 \textbf{x}_t/2$ \cite{Gardiner:2004}:
\begin{equation}
	\partial_t\textbf{x}_t = \textbf{A}(\textbf{x},t) \textbf{x}_t - \frac{1}{2}[\textbf{B}(\textbf{x},t)]^2 \textbf{x}_t + \textbf{B} (\textbf{x},t) \textbf{x}_t \, \xi(t). \label{eqn:SDEStratform}
\end{equation}
Since Stratonovich SDEs satisfy the rules of deterministic calculus, we choose not to use the differential shorthand, which also allows Ito and Stratonovich SDEs to be quickly distinguished. Nevertheless, Eq.~(\ref{eqn:SDEStratform}) should also be strictly interpreted as an integral equation. 

Let us return to the stochastic master equation for the conditional evolution of the unnormalised density operator, $\rho_t$. This is commonly written in the Ito form
\begin{equation}
	d\rho_t = \sL \rho_t dt + \sqrt{\eta} \sA[L e^{i\theta}] \rho_t dy_\theta(t), \label{eqn:itooptimalfilter}
\end{equation}
where $\sL \rho_t = -i[H,\rho_t] + \sD[L] \rho_t$, $\sD[Z] \rho_t = Z \rho_t Z^\dag - (Z^\dag Z \rho_t + \rho_t Z^\dag Z)/2$, $\sA[Z] \rho_t = Z \rho_t+ \rho_t Z^\dag$ and $dy_\theta(t) = \sqrt{\eta} \, \langle L \exp(i\theta) + L^\dag \exp(-i\theta) \rangle_t dt + dw(t)$. Since $dy_\theta(t)^2 = dt$ and $\sD[Z] \rho_t$ and $\sA[L e^{i\theta}]\rho_t$ are linear superoperators, Eq.~(\ref{eqn:itooptimalfilter}) is a linear SDE of the form (\ref{eqn:SDEitoform}). The Stratonovich version is therefore of the same structure as Eq.~(\ref{eqn:SDEStratform}):
\begin{align}
	\partial_t \rho_t 	&= \sL \rho_t  + \sqrt{\eta} \sA[L e^{i\theta}] \rho_t \,  y_\theta(t) - \frac{\eta}{2}\sA^2[L e^{i\theta}] \rho_t,   \label{eqn:system_app}
\end{align}
where $\sA^2[Z]\rho_t= Z (\sA[Z] \rho_t) + (\sA[Z] \rho_t) Z^\dag$.

\subsection{II. No-knowledge measurements and convergence} \label{sec_convergence}
In this section we prove that the quantum filter does not in general converge to the underlying system state if the system is undergoing a no-knowledge homodyne measurement. As stated in Eq.~(5) of the main text, when the system undergoes no-knowledge monitoring (i.e. homodyne detection at an angle $\theta = \pi/2$), the dynamics of the underlying conditional state are given by
\begin{align}
	\partial_t \rho_t 	&= -i\left[ H -  L y_{\pi/2}(t), \rho_t \right].  \label{eqn:repnoknowmasteqn}
\end{align}
We also assume that an observer makes some optimal Bayesian estimate of the system state, $\pi_t$, conditioned on the measurement signal $y_{\pi/2}$. For a no-knowledge measurement, both $\rho_t$ and $\pi_t$ obey Eq.~(\ref{eqn:repnoknowmasteqn}). However, in general the initial conditions differ (i.e. $\pi_0 \ne \rho_0$). 

Define $\Delta_t \equiv \rho_t - \pi_t$, which clearly satisfies Eq.~(\ref{eqn:repnoknowmasteqn}), and will therefore retain its initial normalisation. We now quantify the difference between the system and filter via the Frobenius distance $|| \Delta_t || \equiv \sqrt{\Tr[\Delta_t^2]}$. Due to the form of Eq.~(\ref{eqn:repnoknowmasteqn}), the Frobenius distance is constant in time: 
\begin{align}
	\partial_t || \Delta_t || & = \frac{\Tr\left[ \Delta_t (\partial_t \Delta_t ) + (\partial_t \Delta_t ) \Delta_t \right]}{2|| \Delta_t||} \notag \\
					& = \frac{ -i\Tr \left[ \Delta_t [H - L \, y_{\pi/2}(t), \Delta_t] \right]  }{ ||\Delta_t || } \nn \\
					& = 0. 
\end{align}
Thus $\rho_t$ and $\pi_t$ remain the same distance apart from each other for all time. This shows that under no-knowledge monitoring, it is impossible for an experimenter to refine their estimate of the system state.

\subsection{III. Equations of motion for a qubit undergoing dephasing}
Consider the Stratonovich stochastic master equations 
\begin{subequations}
\label{qubit_SME}
\begin{align}
	\partial_t \rho_t 	&= -i [ \Omega \sigma_x, \rho_t ] + \gamma \sD[\sigma_z]\rho_t  + \sqrt{\gamma} \sA[\sigma_z e^{i \theta}]\rho_t y_\theta(t) \notag \\
				& - \frac{\gamma}{2} \sA^2[\sigma_z e^{i \theta}] \rho_t, \label{2_sys}\\
	\partial_t \pi_t 	&= -i [ \Omega \sigma_x, \pi_t ] + \gamma \sD[\sigma_z]\pi_t + \sqrt{\gamma} \sA[\sigma_z e^{i \theta}] \pi_t y_\theta(t) \notag \\
				& -\frac{\gamma}{2}\sA^2[\sigma_z e^{i \theta}] \pi_t \label{2_filt}
\end{align}
\end{subequations}
which correspond to the physical setup depicted in Fig.~1(a) of the main text. Although equations~(\ref{2_sys}) and (\ref{2_filt}) look similar, it is important to recognise that they are coupled via the same measurement signal 
\begin{equation}
	y_\theta(t) = 2 \sqrt{\gamma} \cos \theta \left(\Tr[ \sigma_z \rho_t ]/ \Tr[\rho_t]\right) + \xi(t). \label{eqn:meas_signal}
\end{equation}
For a qubit, the density matrix for the underlying system takes the simple form
\begin{equation}
	\rho_t = \frac{1}{2}\left( I + x_\rho(t) \sigma_x + y_\rho(t) \sigma_y + z_\rho(t) \sigma_z \right),
\end{equation}
where, for example, $x_\rho(t) = \Tr[\sigma_x \rho_t]/\Tr[\rho_t]$, which implies that $(x_\rho(t), y_\rho(t), z_\rho(t))$ are the co-ordinates defining the Bloch vector. Similarly, $\pi_t = (I + x_\pi(t) \sigma_x + y_\pi(t) \sigma_y + z_\pi(t) \sigma_z)/2$. The equations of motion (\ref{qubit_SME}) therefore reduce to the following set of Stratonovich SDEs:
\begin{subequations}
\begin{align}
	dx_\rho	&= 2 \sqrt{\gamma} \left( y_\rho \sin \theta - x_\rho z_\rho \cos\theta \right) y_\theta(t), \\
	dy_\rho	&= - \Omega z_\rho - 2 \sqrt{\gamma} \left( x_\rho \sin \theta + y_\rho z_\rho \cos\theta \right) y_\theta(t), \\
	dz_\rho	&= \Omega y_\rho + 2 \sqrt{\gamma}(1-z_\rho^2) \cos\theta y_\theta(t), \\
	dx_\pi	&= 2 \sqrt{\gamma} \left( y_\pi \sin \theta - x_\pi z_\pi \cos\theta \right) y_\theta(t), \\
	dy_\pi	&= - \Omega z_\pi - 2 \sqrt{\gamma} \left( x_\pi \sin \theta + y_\pi z_\pi \cos\theta \right) y_\theta(t), \\
	dz_\pi	&= \Omega y_\pi + 2 \sqrt{\gamma}(1-z_\pi^2) \cos\theta y_\theta(t).
\end{align}
\end{subequations}

\subsection{IV. No-knowledge measurement and feedback for photodetection} \label{apx:noknowjumpmast}
Consider again the open quantum system described by Eq.~(1) of the main text. By directly measuring the number of reservoir quanta, the system dynamics can be conditioned according to the stochastic master equation \cite{Wiseman:2010}
\begin{align}
	d\omega_t &= -i[H,\omega_t] dt -  \frac{1}{2}\sA[L^\dag L] \omega_t  dt \notag \\
			& + (L \omega_t L^\dag - \omega_t) dj(t), \label{eqn:jumpeqn}
\end{align}
where $\omega_t$ is the conditional density operator and $j(t)$ is the measurement record, which is a Poissonian process with an average jump rate of $\langle L^\dag L \rangle$. Since this stochastic master equation commonly describes the direct detection of photons emitted from a system, we call such monitoring \emph{photodetection}. 

Knowledge about the system is contained in the rate at which jumps occur. However, when $L=U$ for unitary $U$ (i.e. $U^\dag U = U U^\dag = 1$), the jump rate is always unity, and thus the measurement signal gives no-knowledge of the system dynamics. In this case, Eq.~(\ref{eqn:jumpeqn}) reduces to
\begin{equation}
	d\omega_t = -i[H,\omega_t] dt - (U \omega_t U^\dag - \omega_t) dj(t). \label{eqn:noknowjumpeqn}
\end{equation}
As for the homodyne case, the underlying system state $\omega_t$ and the quantum filter $\tilde{\omega}_t$ will never converge under no-knowledge photodetection. This can be shown explicitly by examining $\Delta_t = \omega_t - \tilde{\omega}_t$, which for $L = U$ satisfies Eq.~(\ref{eqn:noknowjumpeqn}). The evolution of the Frobenius distance is therefore (\emph{c.f.} Sec.~\ref{sec_convergence})
\begin{align}
	d || \Delta_t || 	&= \frac{ -i\Tr \left[ \Delta_t [H, \Delta_t] \right]  }{ ||\Delta_t || } dt \notag \\
				& + \left( || U \Delta_t U^\dag || - ||\Delta_t ||\right) dj(t) \nn \\
 				&= 0,
\end{align}
implying that $\omega_t$ and $\tilde{\omega}_t$ remain an equal distance apart for all time.

Under the evolution (\ref{eqn:noknowjumpeqn}), decoherence can be entirely removed from the system by simply applying the unitary operator $U^\dag$ to the system whenever a jump occurs. For after each jump, the state then becomes $\omega_{t + d t} = U^\dag (U\omega_t U^\dag) U = \omega_t$, implying that only the coherent evolution $d\omega_t = -i[H,\omega_t]dt$ remains.

\end{document}